# The impact of water clouds on the prospective emission spectrum of Teegarden's Star b as observed by LIFE

Ryan Boukrouche [1,2] Rodrigo Caballero [1] and Markus Janson [2]

[1] Department of Meteorology, Stockholm University, Sweden
[2] Department of Astronomy, Stockholm University, Sweden

## ABSTRACT

Non-transiting terrestrial planets will be accessible by upcoming observatories of which LIFE is an example. Planet b orbiting Teegarden's Star is one of the optimal targets for such missions.

We use a one-dimensional atmospheric model with real-gas radiation, a multi-species pseudo-adiabatic convection-condensation scheme, and a water cloud scheme, to estimate the impact of the cloud coverage on the emission spectrum of the target, as well as to assess how sensitive LIFE could be to changes in outgoing flux caused by these clouds.

Though the emergent flux decreases with a higher cloud coverage, it does not decrease by more than one order of magnitude as the coverage increases from 0% to 90%. This allows LIFE to retain a high sensitivity to the cloud cover fraction for wavelengths longer than 7 microns. In this spectral range, with at least 1 bar of $N_2$, LIFE is able to distinguish cloud cover fractions as small as 10% given an integration time of 24 hours, and yields much better precision with a week-long integration. An integration time of one week also allows the resolution of local variations in spectral flux, which can lead to an easier molecule identification. This ability remains if the planet is a $CO_2$-dominated Venus analog. Partial pressures of $N_2$ lower than 1 bar may create a degeneracy with the cloud cover fraction.

LIFE shows promising potential for detecting and characterizing atmospheres even with a high cloud coverage, and retaining a fine sensitivity to relatively small differences in cloud cover fractions.

## 1. INTRODUCTION

LIFE (Quanz et al. 2022) is a candidate direct imaging mission whose science priorities conform to those detailed in the Voyage 2050 long-term plan in the ESA Science Programme. It will attempt to constrain $\eta_{\rm hab}$ and $\eta_{\rm life}$, the occurrence rates of habitability and life on nearby exoplanetary systems, including terrestrial planets orbiting Sun-like stars. It will use nulling interferometry to block the light from the host stars and spatially resolve rocky exoplanets in mid-infrared thermal emission.

The LIFE community has identified Teegarden's Star as one of its golden targets, as it is a nearby ultra-cool M dwarf about 3.8 parsecs away and orbited by two planets discovered through radial velocity (Zechmeister et al. 2019). A third planet was detected by Dreizler et al. (2024). There is as of yet no sign of any of them transiting, which makes them only accessible to direct imaging facilities. Zechmeister et al. (2019) estimated their masses at $1.05^{+0.13}_{-0.12}\,M_{\rm Earth}$ for planet b and $1.11^{+0.16}_{-0.15}\,M_{\rm Earth}$ for planet c. Using a mass-radius relation based on the Preliminary Reference Earth Model (PREM) (Zeng et al. 2016), they showed that assuming a roughly rocky bulk composition leads to radii close to the radius of the Earth. Dreizler et al. (2024) estimated slightly different masses, $1.16^{+0.12}_{-0.11}\,M_{\rm Earth}$ for planet b, $1.05^{+0.14}_{-0.13}\,M_{\rm Earth}$ for planet c, and $0.82^{+0.17}_{-0.17}\,M_{\rm Earth}$ for planet d.

The Teegarden planets are prime candidates for habitability not only because their orbits are within their habitable zones, but also due to the low activity of their host star (Dreizler et al. 2024). Atri & Mogan (2021) suggested that M4 to M10 stars are the least likely to erode secondary atmospheres by stellar flares and XUV irradiation. Teegarden's Star is an M7, which increases the likelihood that planet b retained a secondary atmosphere.

Corresponding author: Ryan Boukrouche
ryan.boukrouche@astro.su.se

Stassun et al. (2019) investigated the habitability of planets b and c, suggesting that their orbits might allow for the presence of liquid water at intervals of latitude that vary depending on the semi-major axis and the efficiency of atmospheric heat redistribution around the planet. This approach lacked realistic representations of gas and cloud radiative transfer and convection.

The present work aims to explore the impact of different cloud configurations on the emission spectra that LIFE might obtain from an $N_2$-$H_2O$ atmosphere with the properties of Teegarden's Star b. We use for this purpose spectrally resolved gas and cloud radiation modelling, as a first step of a broader study that will include more complex compositions and three-dimensional dynamics in subsequent works.

The radiation model is coupled to an idealized column model which enables the exploration of different parameter spaces. We sweep over the fraction of cloud coverage, the surface temperature, and the partial pressure of nitrogen. We compute the resulting spectral thermal emission in each case and use it to estimate the performance of LIFE under different scenarios.

In Sect. 2 we describe the tools used for the modelling and the parameters used. We present results in Sect. 3. We discuss the significance of the results for future observations in Sect. 4 and conclude in Sect. 5.

## 2. METHODS

SOCRATES (Edwards & Slingo 1996) is a real-gas radiation code managed by the UK Met Office. We use it to model atmospheric columns of given compositions with real-gas radiation using the correlated-$k$ method (Goody et al. 1989; Amundsen et al. 2017). The longwave and shortwave spectral files used include 400 bands from 1 $cm^{-1}$ to 35000 $cm^{-1}$ with band widths decreasing toward shorter waves. It includes eight species: $H_2O$, $H_2$, $CO_2$, CO, $CH_4$, $N_2$, $O_3$, and $O_2$. It also has the following self- and foreign-broadened continua: $H_2O$-$H_2O$, $H_2$-$CH_4$, $H_2$-$H_2$, $N_2$-$H_2$, $N_2$-$N_2$, $N_2$-$H_2O$, $O_2$-$CO_2$, $O_2$-$N_2$, $O_2$-$O_2$, $CO_2$-$CO_2$, $CO_2$-$H_2$, and $CO_2$-$CH_4$. The continuum absorption data are obtained from HITRAN 2016 (Gordon et al. 2017), except for the self-broadened continuum of water vapor which was taken from https://github.com/DavidSAmundsen/socrates_tools/tree/master/continuum/h2o. The line absorption data for $H_2O$ and $N_2$ are taken from the ExoMol POKAZATEL (Polyansky et al. 2018) and WCCRMT (Western et al. 2018) line lists respectively.

Here we focus on $H_2O$ and $N_2$ to simplify the interpretation of the outputs of this preliminary study. We assume an $N_2$-dominated atmosphere in a temperate climate. In this context, the main condensible species is water vapor.

Since Teegarden's Star does not yet have a detailed spectrum for radiation calculations, we use the stellar spectrum of TRAPPIST-1 taken from the BT-Settl Model grid of theoretical spectra hosted by the Spanish Virtual Observatory. It is a good proxy because the two stars are of neighboring types and have similar parameters as seen on Table 1, although a key difference is that TRAPPIST-1 is moderately active (Gillon et al. 2017) whereas Teegarden's Star is relatively low in activity (Dreizler et al. 2024).

| Star | TRAPPIST-1 | Teegarden's Star |
|---|---|---|
| Spectral type | M8V | M7V |
| Radius [$R_\odot$] | 0.119 | 0.118 |
| Mass [$M_\odot$] | 0.0898 | 0.0945 |
| Effective temperature [K] | 2566 ± 26 | 2790 ± 157 |
| Source | Agol et al. (2021) | Stassun et al. (2019) |

**Table 1.** Stellar parameters of TRAPPIST-1 and Teegarden's Star.

We use the multi-condensible adiabatic lapse rate generator described by Graham et al. (2021) to generate consistent fully-convective pseudo-adiabatic profiles with an arbitrary number of condensible species as inputs for SOCRATES. It assumes local thermodynamical chemical equilibrium, which is a commonly-used approximation in Earth simulations (Manabe & Wetherald 1967). Wright (2024) is a recent study that focuses on the topic of local thermodynamical equilibrium (LTE) in exoplanet atmospheres, the limits of its validity, and the NLTE regime particularly in the context of hot Jupiters where it is most salient. The upper atmosphere of terrestrial planets experiences a number of effects that steer the environment toward local disequilibrium. However, LTE can still be a good assumption in the lower atmosphere, which is the focus of this study, where the density is high enough that the time between collisions is shorter than the time taken by natural decay toward lower energy states. In addition, the photospheres of temperate worlds where the optical depth is unity are located in the lower atmosphere where LTE is a good approximation, which is why, as we will see, the modelled spectral thermal emission is dominated by continuum absorption. As we model warmer and more optically thick atmospheres, the photosphere rises such that a larger fraction of the thermal emission would come from lower-density layers where including non-LTE effects would play a more important role.

We also use a simple water cloud scheme that defines layers of cloud wherever water saturates and condensates are formed, above the lifted condensation level and as long as the water vapor partial pressure remains higher than an arbitrary value of $10^{-10}$ Pa. This value serves to have a well-defined cloud top depending on the amount of water present in the column. We tested the impact of setting it to zero, so that the cloud systematically extends to the top of atmosphere. Doing so only yields a change in flux of about 1%, and the top of atmosphere fluxes are almost unchanged. Three parameters are input into the cloud radiation scheme of SOCRATES. The first is the effective radius of the droplets, which here is set to 10 microns, which is a typical value used in models focusing on the Earth, where droplets larger than about 20 microns typically result in drizzle. The impact of different radii and distribution widths should be tackled by a follow-up study, as decreasing the particle size for example would increase the shortwave scattering and the planetary albedo. It would also increase the greenhouse effect of the cloud which would act in opposition to the increase in albedo. The latter is typically less sensitive than the former to droplet radius, however quantifying this requires a more complex microphysical treatment that is outside the scope of the present work. The second parameter is the liquid water mass fraction, which describes the mass of liquid water present as condensate in a given cell after saturation has occurred and which contributes to the cloud. This is the mixing ratio of $H_2O$ condensates computed by the multi-condensible adiabat scheme. The last parameter is the water cloud fraction, which represents the fraction of the surface area of a given cell that is taken over by the cloud. This is an unconstrained variable that we explore here. SOCRATES calculates separate fluxes internally for the cloudy and clear-sky fractions of each vertical layer. The way these fluxes are transferred vertically through the column depends on the cloud overlap method used. We use a random overlap which assumes that the presence of a cloud at a given layer is uncorrelated to the presence of any overlapping clouds at neighboring layers. This yields an intermediate result between a maximum overlap where the cloud cover is minimized and the cloud optical depth is maximized, and a minimum overlap where the inverse is true (Lebrun et al. 2023). Of these three approaches, it is the weakest assumption and the most commonly used one in Earth models.

In order to assess the impact of the cloud on key climate metrics such as the modelled tropospheric radiation limit, we modeled a dayside and a nightside column made out of 1 bar of $N_2$ and an initial water ocean of 260 bar, for each surface temperature in a range between 200 K and 3000 K. The ocean evaporates into the atmosphere depending on the surface conditions. At low enough temperatures for which water vapor is still a trace gas, the surface pressure tends toward the partial pressure of the only other constituent, $N_2$. As the surface warms up and more water evaporates into the atmosphere, the partial pressure of water remains equal to the saturation vapor pressure so that the relative humidity is unity across the column. When the surface temperature is large enough that the saturated profile contains the same amount of water as the initial ocean inventory, the partial pressure of water stops increasing because there is no more ocean water to evaporate, and the bottom layers of the atmosphere become subsaturated upon further surface warming. With an initial water ocean of 260 bar in our setup, the point at which the lowest layer becomes subsaturated happens at a surface temperature of about 600 K. This water vapor feedback is a key mechanism during a runaway greenhouse effect (Kasting 1988; Boukrouche et al. 2021).

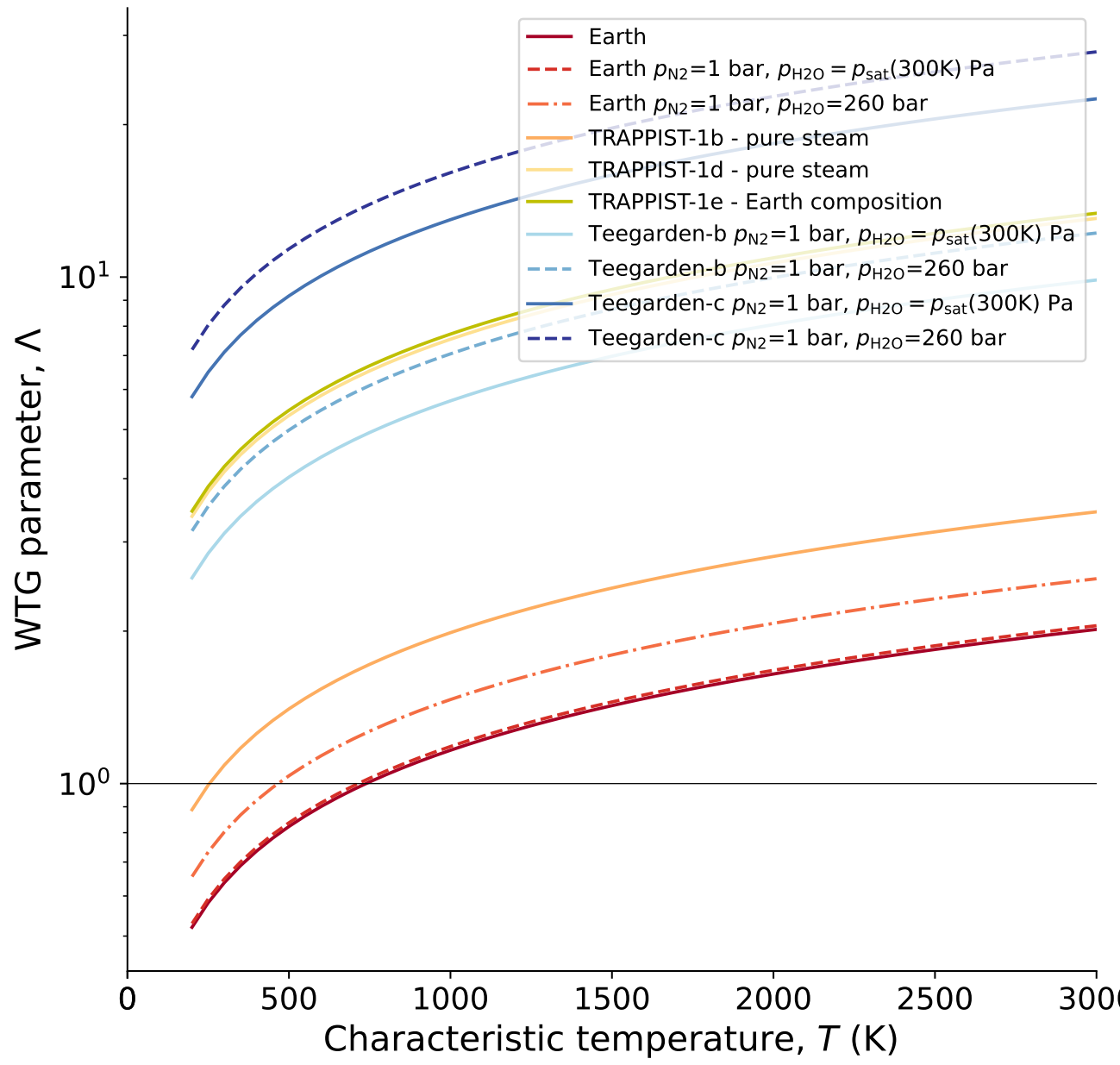

**Figure 1.** Weak temperature gradient (WTG) parameter $\Lambda$ as a function of the characteristic temperature of different atmospheres. If $\Lambda$ is smaller than unity, it indicates an atmosphere that can support significant horizontal temperature gradients away from the equator. The further above unity $\Lambda$ is, the deeper the atmosphere is in a WTG regime.

The inclusion of 1 bar of $N_2$ is a strong assumption since it is currently not possible to directly constrain its abundance in exoplanetary atmospheres due to its lack of spectral features. Hall et al. (2023) suggested that with the advent of future direct imaging missions, there may be ways to use the pressure broadening, collision-induced absorption, and the absence of spectral features of $N_2$ in specific regions of the spectrum to rule out the presence of a non-condensible background gas other than $N_2$ and put indirect constraints on its abundance. In this work, we included a parameter sweep on the abundance of $N_2$ in order to estimate its impact on the results.

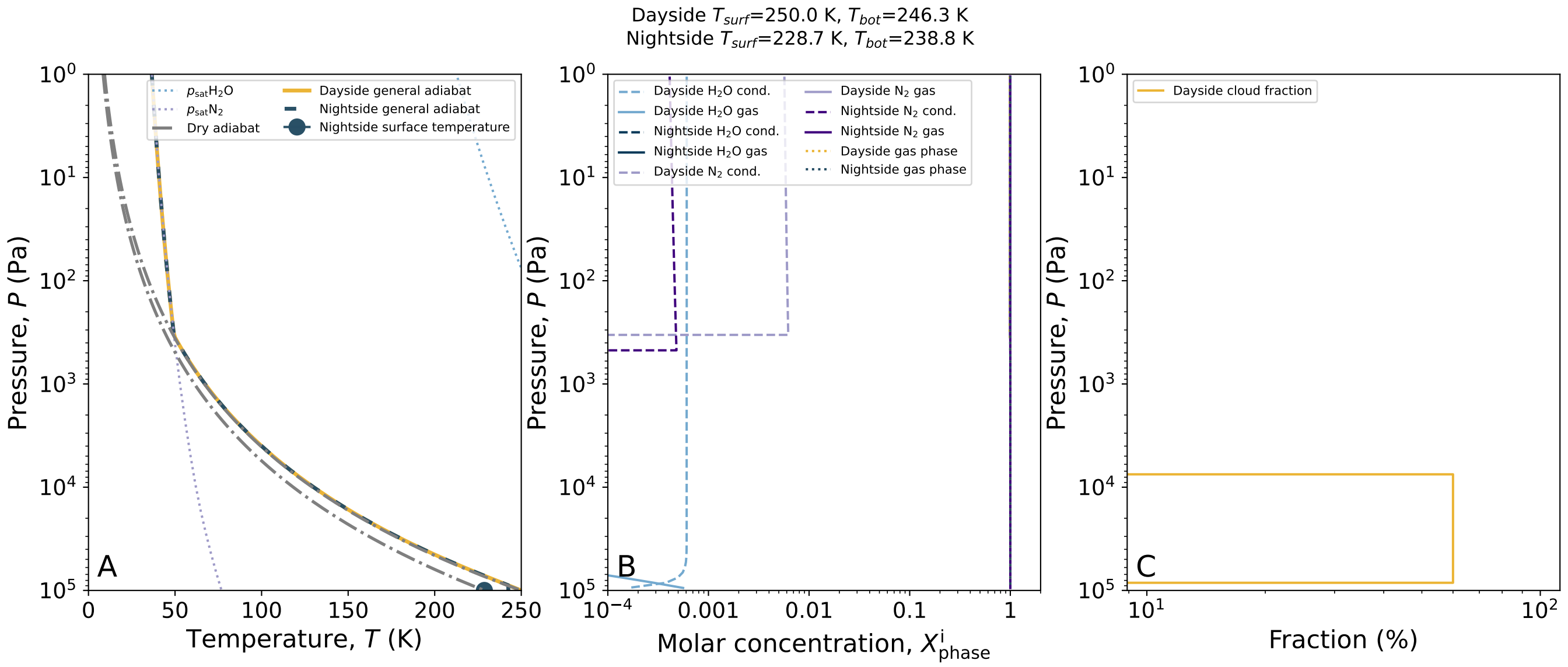

**Figure 2.** **(A)** Temperature profiles at a dayside surface temperature of 250 K. **(B)** Mixing ratios of gaseous and condensed phases. **(C)** Cloud coverage profile with a cloud cover fraction set to 60%.

The stated goal of LIFE to constrain the occurrence rate of habitability may also help determine the occurrence rate of Earth-like climates and Venus-like climates. The ability to make this distinction will be critical, since the two have reached very different end states despite having relatively similar bulk properties, and defining the inner edge of a habitable zone is heavily model-dependent. There have been preliminary attempts to estimate the occurrence rates of Venus analogs based on current catalogs (Kane et al. 2014), however in this study, we cannot discount the possibility that the Teegarden planets have experienced a runaway greenhouse effect and are currently in a Venus-like state. Therefore, we included a sweep on the cloud fraction with a roughly Venus-like composition of 3.3 bars of $N_2$, 89.7 bar of $CO_2$, and 7 bar of evaporated water vapor, for a total surface pressure of 100 bar.

Zechmeister et al. (2019) estimated the age of Teegarden's Star by various methods, and bracketed it between 4 and 10 Gyr. Due to their close orbital distances, the planets around Teegarden's Star have likely had time to complete their synchronization long ago and are now probably tidally locked (Barnes 2017). Chiefly because of this, they should also be in a deep weak temperature gradient regime (WTG) (Pierrehumbert & Hammond 2019), as estimated by Figure 1. This figure shows the WTG parameter $\Lambda$, which represents the proximity of an atmosphere to a WTG regime, for different atmospheric compositions as a function of the characteristic atmospheric temperature. $\Lambda$ is the ratio of the Rossby deformation radius to the planetary radius as evaluated by Equation 1 (Pierrehumbert & Ding 2016).

$$\Lambda = \frac{\sqrt{RT}}{\Omega R_{\mathrm{p}}}, \tag{1}$$

with $R$ the specific gas constant of the dominant species in the atmosphere, $T$ the characteristic temperature of the atmosphere, $\Omega$ the angular velocity of the planet, and $R_{\mathrm{p}}$ the radius of the planet.

As $\Omega$ is small for tidally locked planets, the Rossby deformation radius becomes larger than the planet radius, which means that the Coriolis force is too small to maintain significant horizontal temperature gradients. The atmosphere of the Earth with its current chemical composition is represented by the solid red curve. It would need a characteristic temperature of about 750 K to fall into WTG globally. A characteristic temperature of 500 K would be enough if the Earth's oceans evaporated into the atmosphere over the course of a runaway greenhouse effect, which may warm the surface of the planet to temperatures beyond 2000 K in this case (Kasting et al. 1993; Kopparapu et al. 2013; Boukrouche et al. 2021), pushing it deeper into WTG. At the equator, the Coriolis force is zero, so the Earth is in a WTG regime there regardless of its angular velocity or atmospheric composition. The planets around TRAPPIST-1 and Teegarden's Star would likely be in a global WTG regime regardless of the characteristic temperature of their atmospheres, with the exception of TRAPPIST-1b, which could support significant horizontal temperature gradients given a cold enough characteristic temperature or the absence of an atmosphere.

The nightside column is partly determined by the dayside in this idealized one-dimensional framework, with a method similar to the one used by Yang & Abbot (2014). We first compute the dayside temperature profile, outgoing planetary flux $\mathrm{OPR}_{\mathrm{day}}$, and incoming stellar radiation ISR given a surface temperature $T_{\mathrm{s,day}}$. Because the target planet is in a deep WTG regime, we can assume that the nightside has a similar temperature profile to the dayside. A corollary of this is that we can also assume a similar downward thermal flux profile. We use the latter evaluated at the bottom pressure layer to compute the nightside surface temperature $T_{\mathrm{s,night}} = \left(\frac{F_{\mathrm{s}\downarrow,\mathrm{day}}}{\sigma}\right)^{\frac{1}{4}}$ assuming the surface is a blackbody. We do not use timestepping in this framework. Timestepping and the inclusion of boundary layer turbulence would form a stable inversion linking the nightside surface temperature to the overlying profile. For the purposes of this study, we neglect the inversion and leave the nightside surface separate from the atmospheric profile, as seen on panel A of Figure 2. Panel B shows the mixing ratios of the different phases, featuring about 0.06% of water at the surface and 99.944% of molecular nitrogen. Water condenses at the boundary layer, where the gas phase is replaced by the condensed phase. Without the inversion expected from stellar heating, $N_2$ also condenses around 400 Pa. Panel C shows the cloud cover fraction. Above about 75 mbar, the partial pressure of water drops below $10^{-10}$ Pa, which is the arbitrary threshold we chose to allow cloud formation.

Figure 5 of Hammond & Lewis (2021) describes the overturning circulation on a tidally-locked planet and shows how the substellar point features warm, wet air ascending, precipitating over the terminator, and descending as cold, dry air on the antistellar point. This is a similar picture to the Earth pattern when the equator is analogous to the substellar point and the poles are analogous with the antistellar point. This pattern implies that if cloud formation happens, it is more likely to happen on the dayside rather than the nightside, because the latter would be extremely dry. This pattern applies to a temperate scenario. Indeed, if we assume that the planet is beyond its runaway greenhouse instellation threshold, the dayside would be too warm to allow water condensation and cloud formation, as shown by Turbet et al. (2021) and Boukrouche (2023). Based on these considerations, we assume that about 90% of the saturated water vapor on the dayside is retained as condensates in the atmosphere, and the remaining 10% is lost through precipitation. On the nightside, we assume that any condensate forming is rained out.

Figure 8 of the same paper shows how the pattern would change in the case of a hot Jupiter with a surface pressure of 220 bar, which is a relevant order of magnitude because a rocky planet that evaporated an Earth ocean of water through a runaway greenhouse effect would reach a surface pressure of at least 260 bar. It suggests that dayside ascent of warm and moist air is still present in this case, however the descent may be shifted closer to the terminator, while the nightside may feature an ascent, though much weaker than on the dayside. This highlights the fact that the scenario of a longitudinal dipole with a dayside featuring moist convection favoring cloud formation and a very dry, clear sky nightside is predicated on the assumption that the surface pressure remains close to present-day Earth.

SOCRATES also outputs spectrally resolved fluxes at each pressure level. Once scaled to the target star's distance, these can be used as inputs for LIFEsim (Quanz et al. 2022), which is a software developed by the instruments team of the LIFE collaboration and used to simulate observations by LIFE using various settings, such as the spectral resolution of the instrument, the integration time, the diameter of a single aperture, and the bounds of the spectral range. The latter two settings are used to define what the LIFE team currently views as baseline, pessimistic, and optimistic scenarios. In this work, we assume a baseline scenario with a 2-meter diameter aperture and a spectral range between 4 and 18.5 $\mu$m.

## 3. RESULTS

Panels A and C of Figure 3 shows the dayside and nightside fluxes corresponding to Figure 2. Panel B is the same figure as Panel A, except that the cloud has been removed. The greenhouse effect of the cloud increases the downward thermal flux near the cloud top around 75 mbar, and its albedo effect increases the upward shortwave flux through scattering.

Figure 4 shows the profiles when the surface reaches 650 K. At this temperature, the initial 260 bar ocean is depleted, meaning that all the liquid water has evaporated into the atmosphere as we keep the air saturated. The lifted condensation level rose above the surface whose pressure is now 261 bar, and the high optical thickness of the deep atmospheric layers makes the nightside and dayside surface temperatures converge. As water vapor turns into water condensates, $N_2$ takes the place of the lost vapor, which yields the increase in $N_2$ gas mixing ratio on panel B. This creates a day-side water cloud base on panel C starting at the lifted condensation level and extending to the top of the atmosphere. Figure 5 shows the relative humidity at 250 K and 650 K. At 250 K the profile is above 100%, which means that the air is supersaturated, up to the lifted condensation level around 330 Pa. The mixing ratio of water is about 560 ppm at the bottom and drops to 8.6 $10^{-45}$ at the LCL. There is a numerical instability at this point, where the difference between the saturation vapor pressure of water and its partial pressure is on the order of $10^{-42}$, and their slopes change slightly differently. At 650 K the water vapor mixing ratio is large enough that the atmosphere is nearly pure steam, and therefore has enough water to keep the entire profile saturated except at the bottom layers, which are warm enough to subsaturate the air.

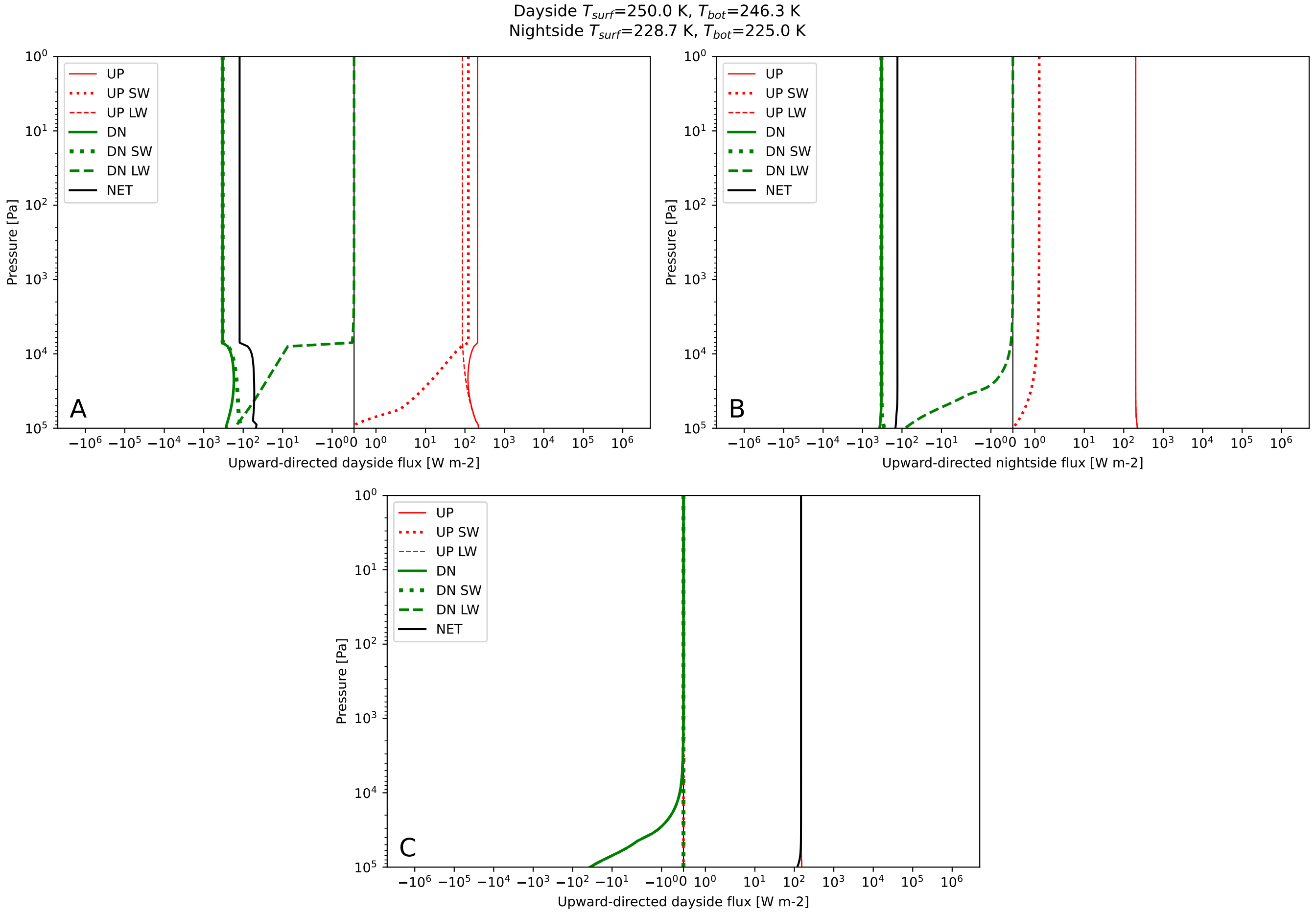

**Figure 3.** **(A, B)** Dayside and **(C)** nightside flux profiles with a dayside surface temperature of 250 K and a cloud fraction of **(A)** 60% and **(B)** 0%.

Figure 6 shows the spectrally-resolved contribution functions at 250 K and 650 K, which quantify the number of watts per square meter that each atmospheric layer contributes to the outgoing planetary radiation, for each spectral bin. The figure also indicates the location of the cloud. The cloud cover fraction does not affect the $\mathcal{CF}_\mathrm{F}$ to any noticeable degree. At 250 K the surface is cold enough that most of the outgoing flux is coming from the cloud layers deeper than 0.4 bar, 0.6 bar within the spectral range of LIFE, and from waves longer than 6 $\mu$m. At 650 K, the bolometric radiating pressure levels rise between 0.3 and 30 mbar. Two of the main water vapor atmospheric windows add a small contribution deeper down and are partially captured by the spectral range of LIFE.

Figure 7 shows the upward photon flux emitted at the top of the atmosphere assuming a surface temperature of 300 K as a function of wavelength, over the same spectral range as before, and for a range of cloud cover fractions. Water vapor absorption can make the emission drop by 2 to 4 orders of magnitude. Increasing the cloud cover fraction decreases the top of atmosphere emission linearly, as illustrated by Figure 8, where the photon flux at selected wavelengths has been plotted as a function of the range of cloud cover fractions displayed, using a linear and logarithmic vertical axis. Decreasing the flux by one order of magnitude requires increasing the cloud cover fraction by more than 90% relative to any given initial value. A significant amplitude could still be observed even if most of the planet is covered by clouds, with the shape of the two water vapor windows included in the LIFE spectral range mostly preserved.

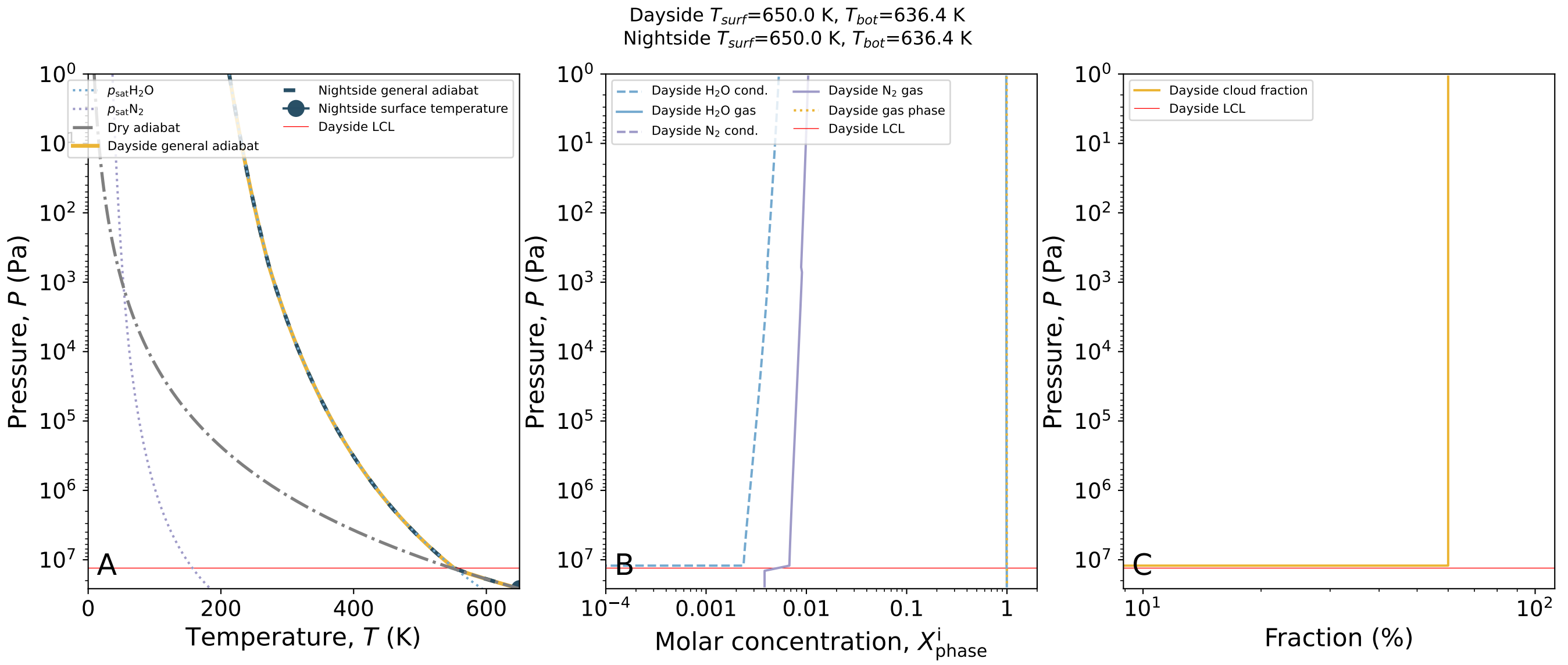

**Figure 4.** **(A)** Temperature profiles at a dayside surface temperature of 650 K. **(B)** Mixing ratios of gaseous and condensed phases. **(C)** Cloud coverage profile with a cloud cover fraction set to 60%. The horizontal red line is the lifted condensation level.

Figure 9 summarizes the outputs from all parameter sweeps performed, over the surface temperature and the cloud cover fraction. The top panel shows the top of atmosphere fluxes. The solid lines are the dayside outgoing planetary radiation $\mathrm{OPR_{day}}$, for which the cloud cover fraction increases from yellow to brown. The green dotted line is $\mathrm{OPR_{night}}$, which is cloud-free. The dash-dotted lines are the absorbed stellar radiation $\mathrm{ASR} = \mathrm{ISR} - \mathrm{F_{\uparrow SW}}$, where ISR is the incoming stellar radiation and $F_{\uparrow SW}$ is the stellar flux scattered upward. For the ASR, the cloud cover fraction increases in shades of blue. The clear-sky case recovers a radiation limit of 278 $\mathrm{W\,m^{-2}}$ which is in line with the values found by previous works (Kopparapu et al. 2013; Hamano et al. 2015; Goldblatt et al. 2013; Boukrouche et al. 2021; Selsis et al. 2023). A planet with an Earth ocean and the orbital parameters of Teegarden b subject to a runaway greenhouse effect would find an equilibrium by warming its surface up to about 2700 K. The overshoot of the radiation limit seen around $T_{\mathrm{surf}} = 400$ K is, as explained by Goldblatt & Watson (2012), due to the presence of $N_2$ which, at low concentrations of water vapor, increases the moist adiabatic lapse rate, which in turn increases the temperature of the radiating pressure levels and raises the emitted flux. As water vapor comes to dominate the atmosphere at higher temperatures, the lapse rate decreases toward the dew point curve of water and the pure steam radiation limit is recovered. As we increase the cloud cover fraction, the OPR decreases linearly as described by Figure 8, and the ASR also decreases linearly due to the albedo effect of the cloud which raises $F_{\uparrow SW}$, making the OPR and the ASR converge. This is also seen on the bottom panel, which shows the planetary albedo as a function of surface temperature. A clear-sky case with a blackbody surface yields a very low albedo. A total cloud cover around Teegarden's Star or TRAPPIST-1 would yield an albedo close to 60%. Around a G-type star like the Sun, the albedo is found to reach a value close to 70%.

Figures 10, 11, and 12 show the results of synthetic observations of Teegarden's Star b with LIFE, computed using LIFEsim. The top panels show the emission spectrum over the spectral range of LIFE and the observed data points outfitted with the calculated error bars and shaded with the 1$\sigma$ confidence areas. The latter are defined within one standard deviation of the noise around the synthetic signal, assuming the noise is normally distributed. The cloud cover is varied over a selected range of fractions. The bottom panels show the statistical significance of the detected difference between the clear-sky case, which serves as a reference, and the other cloudy cases. The significance is calculated with the formula

$$\mathrm{Significance} = \frac{|F_{0\%} - F_{X\%}|}{\frac{F_{0\%}}{\mathrm{SNR}_{0\%}}}, \tag{2}$$

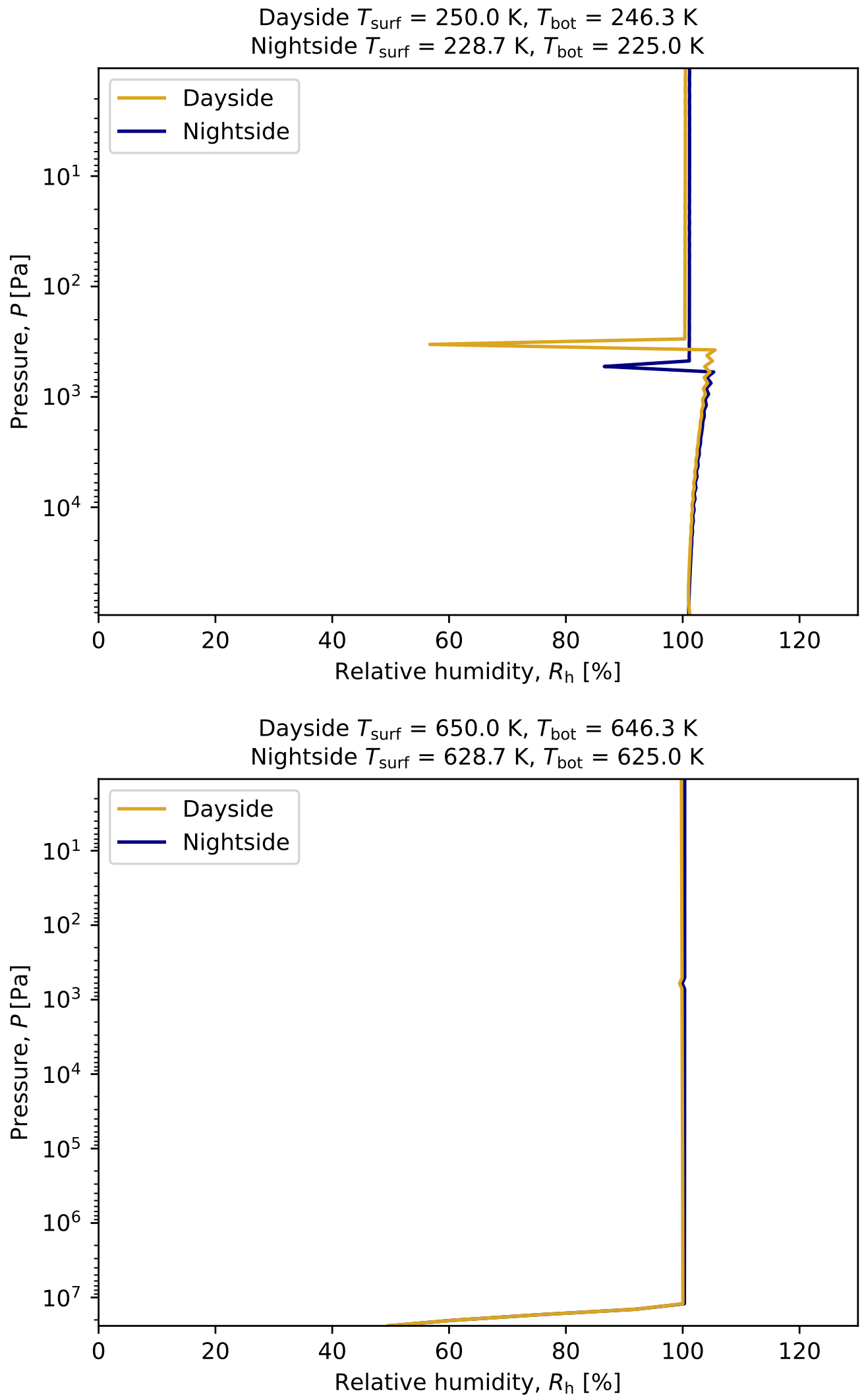

**Figure 5.** Relative humidity profiles for a dayside surface temperature of **(Top)** 250 K and **(Bottom)** 650 K.

where $F_{0\%}$ is the clear-sky flux, $F_{X\%}$ is the flux with a cloud cover fraction of X%, and $SNR_{0\%}$ is the signal to noise ratio associated to the clear-sky flux observation. Figure 10 simulates an observation using a spectral resolution of 50 and an integration time of 24 hours. Error bars are small enough that LIFE can distinguish between cloud cover cases at wavelengths higher than around 7 $\mu$m. The flux decreases sharply below 7 $\mu$m, where the error bars overlay each other. The resulting significance of the distinction between cases peaks around 9.2 $\mu$m and reaches 68. An increase in cloud cover of 10% results in a maximum increase in significance of 6.8 at the peak, relative to the clear-sky case. There is a secondary maximum at 4.8 $\mu$m peaking at 3.18, where an increase in cloud cover of 10% results in an increase in significance of 0.32. The peak seen around 6.3 $\mu$m is due to the central transmission peak inside the absorption region centered around it. Figure 11 increases the integration time to one week. This reduces the error bars and the 1$\sigma$ confidence area so that we are now able to resolve the local variations in flux, again at wavelengths higher than 7 $\mu$m. An increase in cloud cover of 10% results in a maximum increase in significance of 18, reaching 180 for the 0%-100% case. The significance increase for the same ratio is 0.84 at the secondary maximum around 4.8 $\mu$m, and peaks at 8.4. Figure 12 adds 40 Pa of $CO_2$ into the atmosphere, which does not change the picture except for the absorption bands of $CO_2$ around 4.3 and 15 $\mu$m, which effectively brings the fluxes down to the one associated to a total cloud coverage in these regions.

Figure 13 shows a sweep over the partial pressure of $N_2$ from 0.01 bar to 100 bar, with a surface temperature of 300 K, a cloud cover fraction of 90% and the same water vapor scheme as before. This has a large impact on both the sensitivity and the significance. With 0.01 or 0.1 bar of $N_2$, the resulting flux could be difficult to distinguish from a scenario with 1 bar of $N_2$ and a cloud cover fraction of 50 or 60% based on Figure 10. Partial pressures of $N_2$ higher than 1 bar would be difficult to distinguish from each other given the same cloud cover, which also means that the amount of $N_2$ would be much less degenerate with the cloud cover fraction compared to partial pressures of $N_2$ lower than 1 bar.

Figure 14 shows a scenario of a post-runaway atmosphere with 89.7% $CO_2$, 3.3% $N_2$, and 7% $H_2O$. The surface has a Venus-like temperature of 740 K. Both sensitivity and significance are dominated by three main peaks around 8.8, 10, and 11.5 $\mu$m. Although slightly lower, they have the same order of magnitude as the 1 bar $N_2$-dominated case shown by Figure 10. The other spectral regions are dominated by water vapor absorption and the $CO_2$ contribution around 15 $\mu$m.

## 4. DISCUSSION

Increasing the cloud cover fraction on the dayside results in the scenario suggested by Yang & Abbot (2014), which was predicated on the case of a tidally locked Earth-like atmosphere in the habitable zone of M-dwarf stars. In this scenario, tidal-locking may drive intense dayside moist convection, leading to thick tropospheric clouds which shield the planet from a runaway greenhouse effect by scattering a fraction of the incoming stellar radiation to space, effectively pushing the inner edge of its habitable zone starward. This phenomenon may be possible if temperatures are low enough to allow water to condense on the dayside. Then if for instance the star brightens up over time and pushes the inner edge further outward than the planet's orbit, the cloud shielding can take effect. However if the planet started out with at least ten bars of water, as the instellation increases toward its runaway greenhouse radiation limit, infrared absorption from a partial pressure of water vapor that increases due to surface evaporation will warm up the dayside enough to prevent water condensation and thus cloud formation (Turbet et al. 2021; Boukrouche 2023). This, added to a nightside stratospheric cloud formation which compounds the greenhouse effect, would increase the planet's vulnerability to a runaway greenhouse effect. In the aforementioned scenario where dayside condensation is possible, the top-of-atmosphere energy budget of planet b would equilibrate roughly linearly with increasing dayside tropospheric cloud coverage.

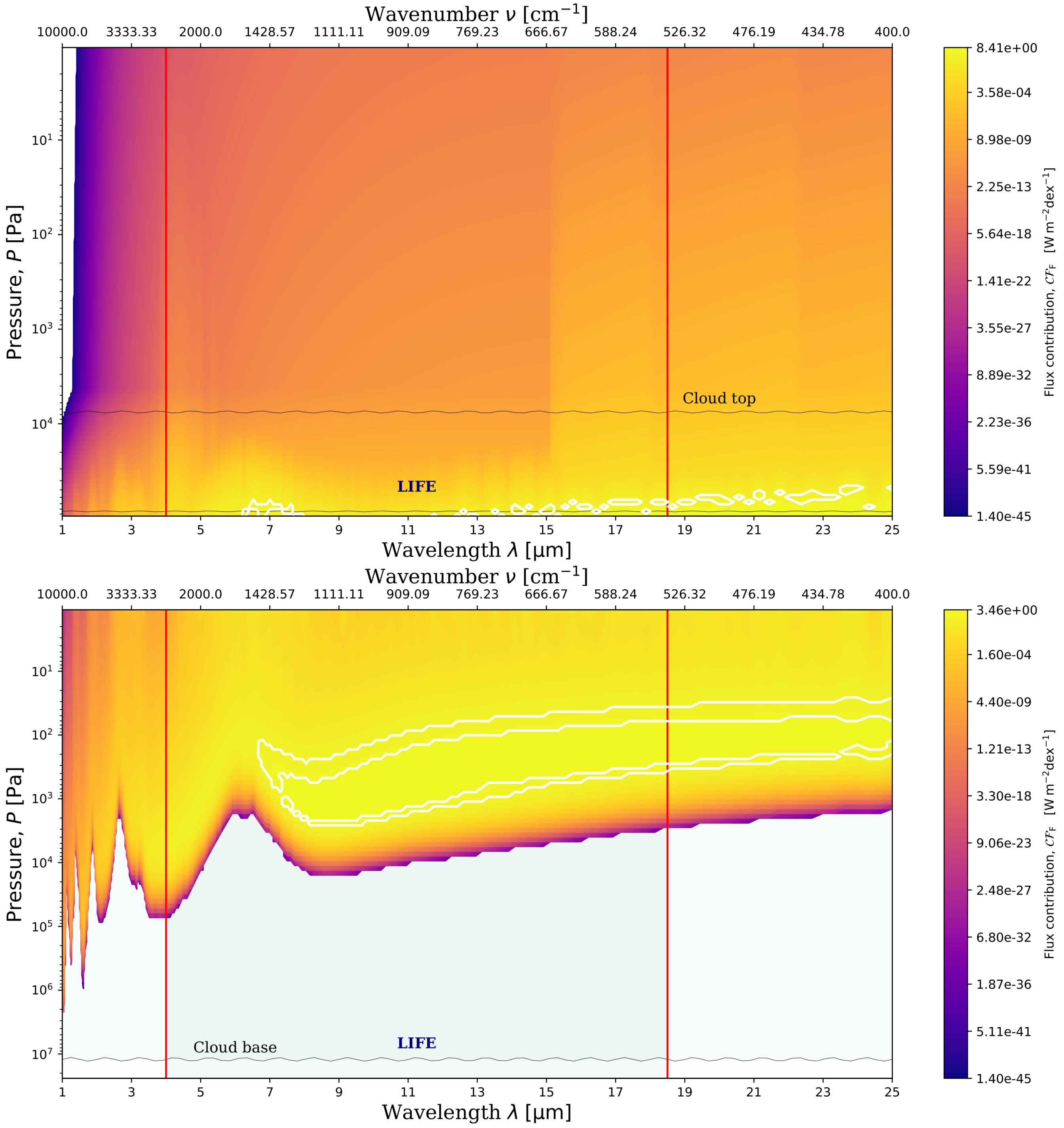

**Figure 6.** Spectral flux contributions over the range [1, 25] $\mu$m for a surface temperature of **(Top)** 250 K and **(Bottom)** 650 K. The white lines delineate the areas where the $\mathcal{CF}_{\mathrm{F}}$ per spectral band is higher than unity. The spectral range of LIFE and the vertical extent of the cloud are overplotted.

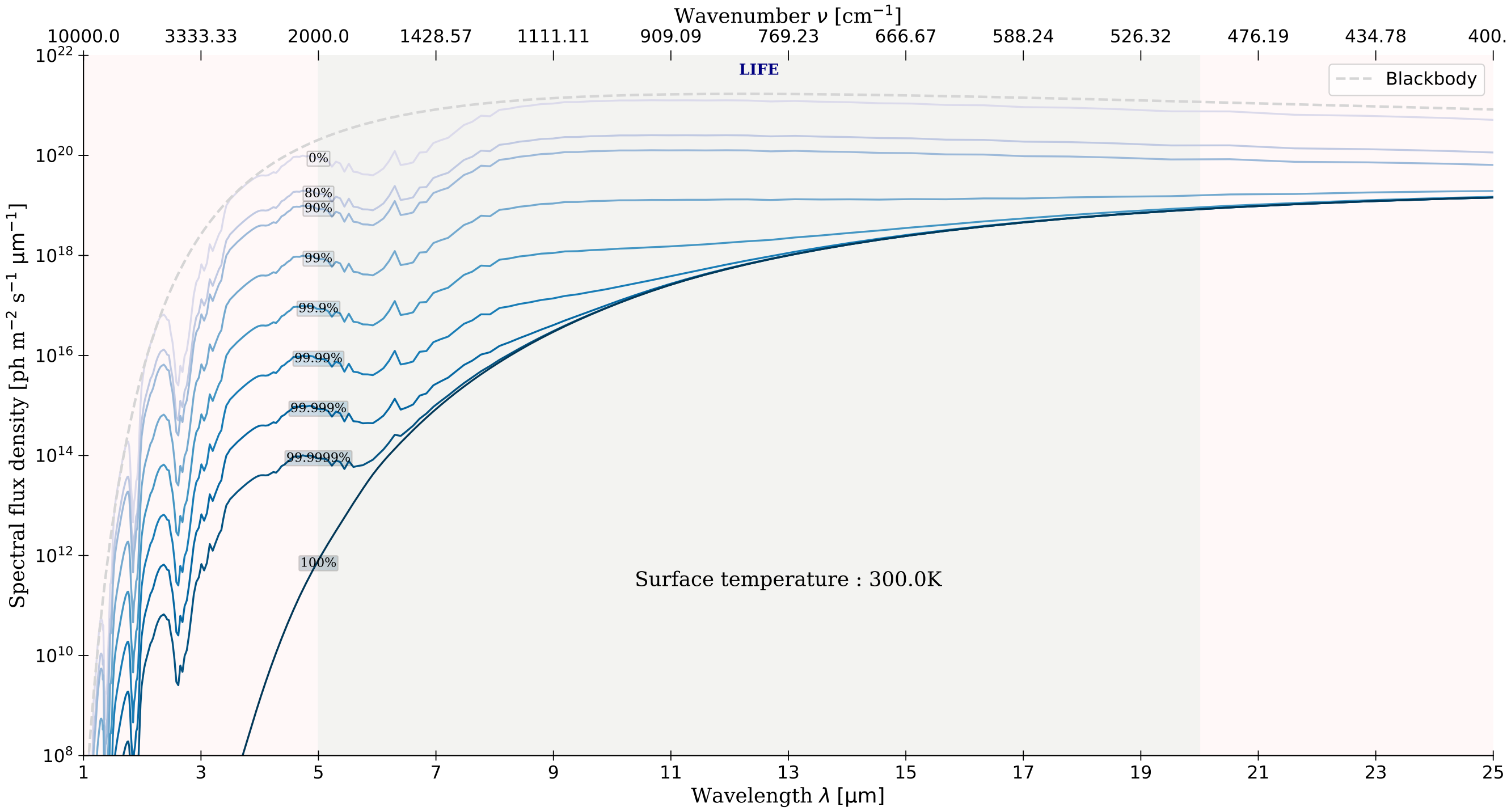

**Figure 7.** Upward photon top-of-atmosphere fluxes as a function of wavelength at a surface temperature of 300 K, for a range of cloud cover fractions. The black dashed curve is the Planck emission of a surface of 300 K and zero albedo.

The planets around Teegarden's Star are within their theoretical habitable zones defined by the range of orbits that permit the existence of liquid water (Kasting et al. 1993; Kopparapu et al. 2013) given an Earth-like atmosphere. In addition, the age of Teegarden's Star estimated by Zechmeister et al. (2019) suggests that the planets have had time to form temperate secondary atmospheres. The goal of LIFE is to observe such potentially habitable planets according to the current primary criterion for habitability, namely the presence of liquid water on their surface. We still have to assume that Teegarden's Star, being an M star, has not fully blasted away the atmospheres of its planets through XUV irradiation and flares. Atri & Mogan (2021) suggested that M4 to M10 stars are the least likely to erode secondary atmospheres by these mechanisms. Given that Teegarden's Star is an M7, it seems possible that planet b was able to retain a secondary atmosphere.

The current baseline spectral range planned for LIFE would have access to two major water vapor windows and a detectable thermal emission despite most of the planet's surface area being covered by water clouds. Given the orbit of planet b around Teegarden's Star, if it has outgassed a significant reservoir of water during its mantle solidification phase or acquired it from later imports from asteroids and comets, water vapor is likely a minor constituent of the atmosphere, as the majority would be condensed on the surface. Even so, assuming a water cycle is active on this planet, a near-total cloud coverage is still possible. The cloud coverage on the Earth is difficult to accurately determine due to its seasonal and daily variability, but it is on average around 60 or 70% (e.g. King et al. (2013)). There is so far no theoretical way to predict this value, or to predict the coverage expected on an exoplanet or even on Earth. Datseris et al. (2022) is a recent attempt to tackle this problem. High clouds formed by rising moist air or low clouds from surface evaporation can spread by detrainment to potentially cover the entire globe, even for very low water partial pressures. In this case, the detected flux might lack any molecular signature. If we assume a cloud coverage lower than 90% however, given the rate at which thermal emission decreases when the cloud fraction increases, it is unlikely that the observed emitted flux would differ from the clear-sky flux by more than one order of magnitude, and spectral signatures would remain accessible.

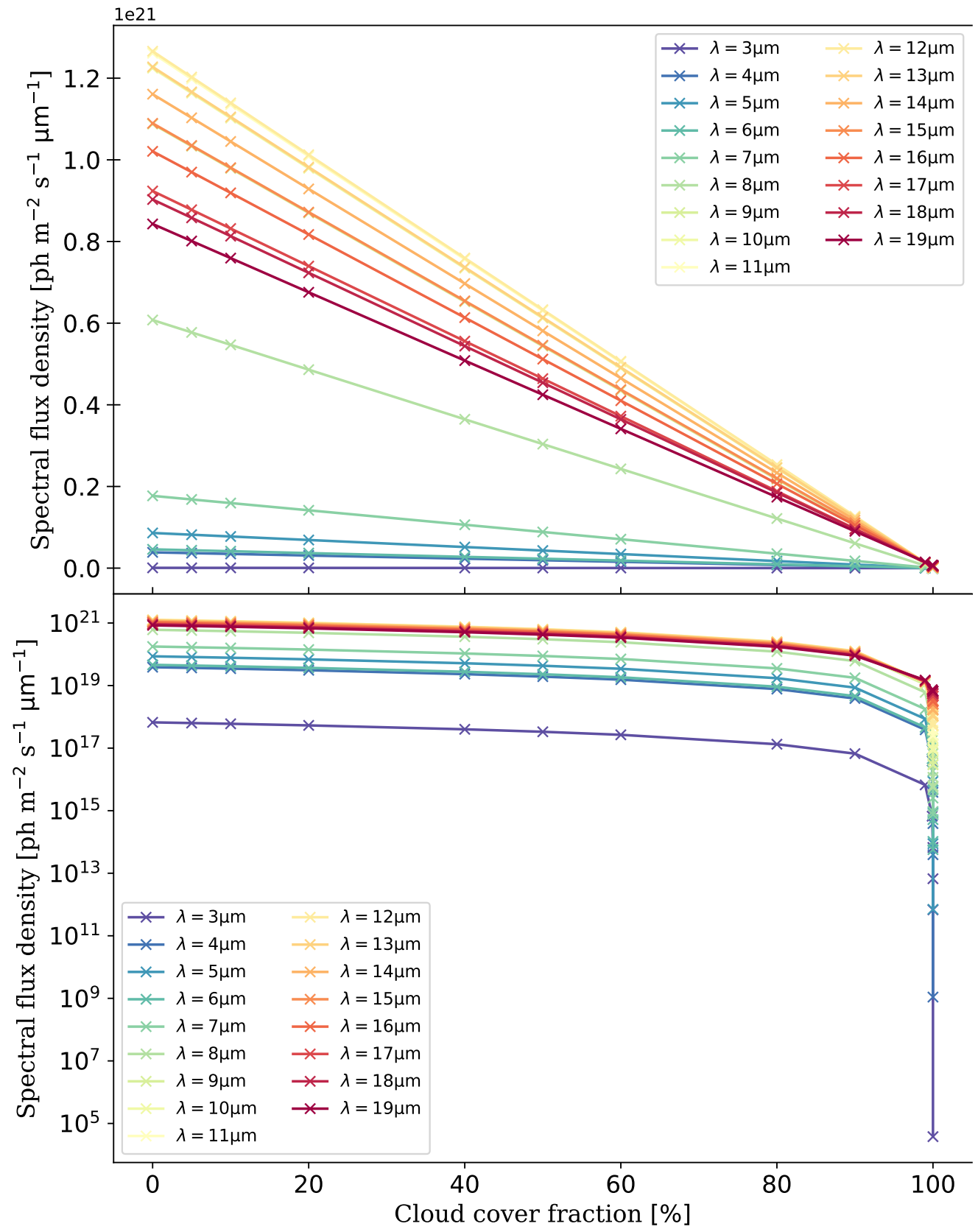

**Figure 8.** Upward photon top-of-atmosphere fluxes for given wavelengths as a function of the cloud cover fraction, with a **(Top)** linear and **(Bottom)** logarithmic vertical scale.

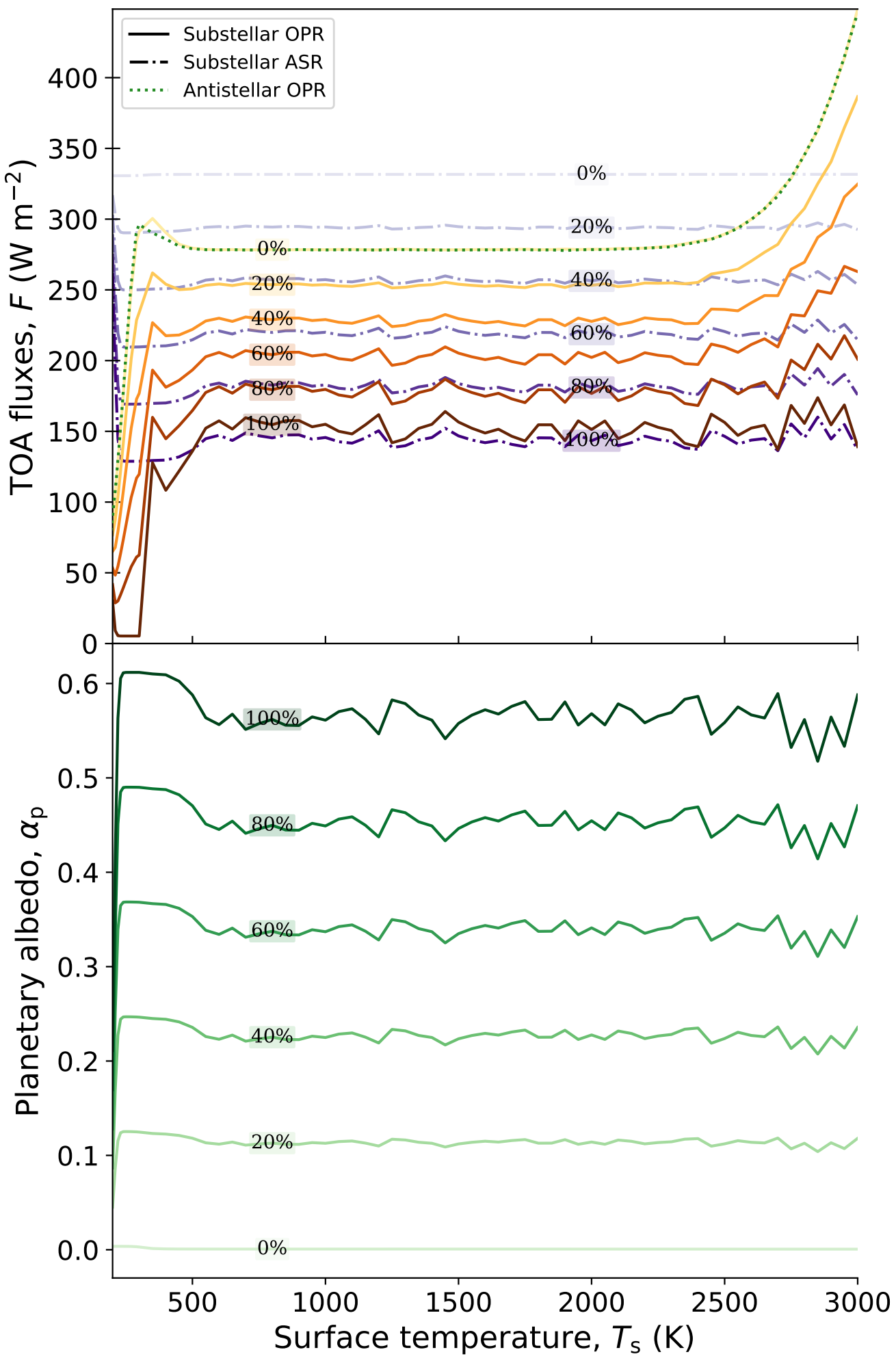

**Figure 9.** **(Top)** Top of atmosphere fluxes as a function of surface temperature. The solid curves from yellow to brown are the dayside outgoing planetary fluxes varying the cloud cover fraction. The green dotted curve is the nightside clear-sky outgoing planetary flux. The dash-dotted curves in shades of blue are the absorbed stellar fluxes varying the cloud cover fraction. **(Bottom)** Planetary albedo as a function of surface temperature for a range of cloud cover fractions.

This is a key advantage of LIFE and other direct imaging missions on transit spectroscopy. The light collected by the James Webb Space Telescope in transmission would be subject to absorption by the horizontal length of a potential cloud, which can be much greater than its vertical extent. The extra absorption can mute the spectral features of species below the cloud top, as described by Komacek et al. (2020), especially for slowly rotating planets, since they are more prone to developing thick dayside cloud decks that might extend to the terminator. As a direct imaging mission, LIFE would collect light that would experience at most absorption by the entire vertical extent of the cloud deck. In addition, the 6.5 mas angular separation of Teegarden b from its parent star is too small to reach a sufficient contrast to detect the planet with JWST, ELT, or any other existing or planned single-dish telescope. Carrión-González et al. (2023) in particular estimated that LIFE will be the only mission able to probe the atmospheres of the Teegarden planets among all other planned ones.

Most of the flux recovered by LIFE would come from the interval starting from 7 microns up to the chosen upper boundary of the spectral range, within or below the cloud layer at temperate conditions. As the atmosphere gets warmer and more opaque, the peak of the contribution remains in the same spectral interval, but rises quickly in the column and could potentially find itself near to or above the cloud top.

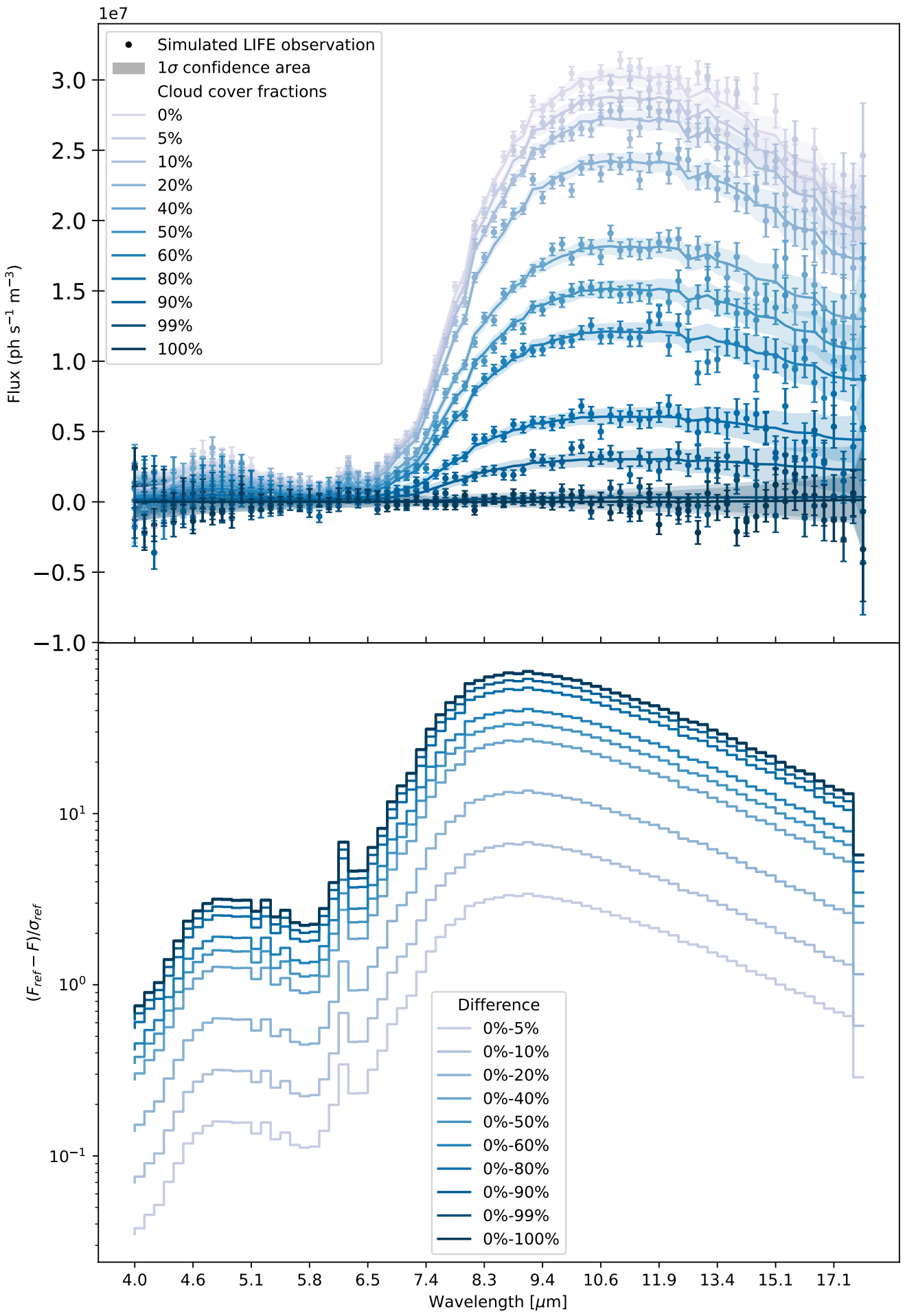

**Figure 10.** LIFEsim simulation of an observation of Teegarden's Star b with a resolution of 50 and an integration time of 24 hours. **(Top)** Upward photon top-of-atmosphere fluxes as a function of wavelength at a surface temperature of 300 K, for a range of cloud cover fractions. The shaded areas represent the $1\sigma$ confidence level. **(Bottom)** Statistical significance of the detected difference between the clear-sky scenario and the cloudy scenarios.

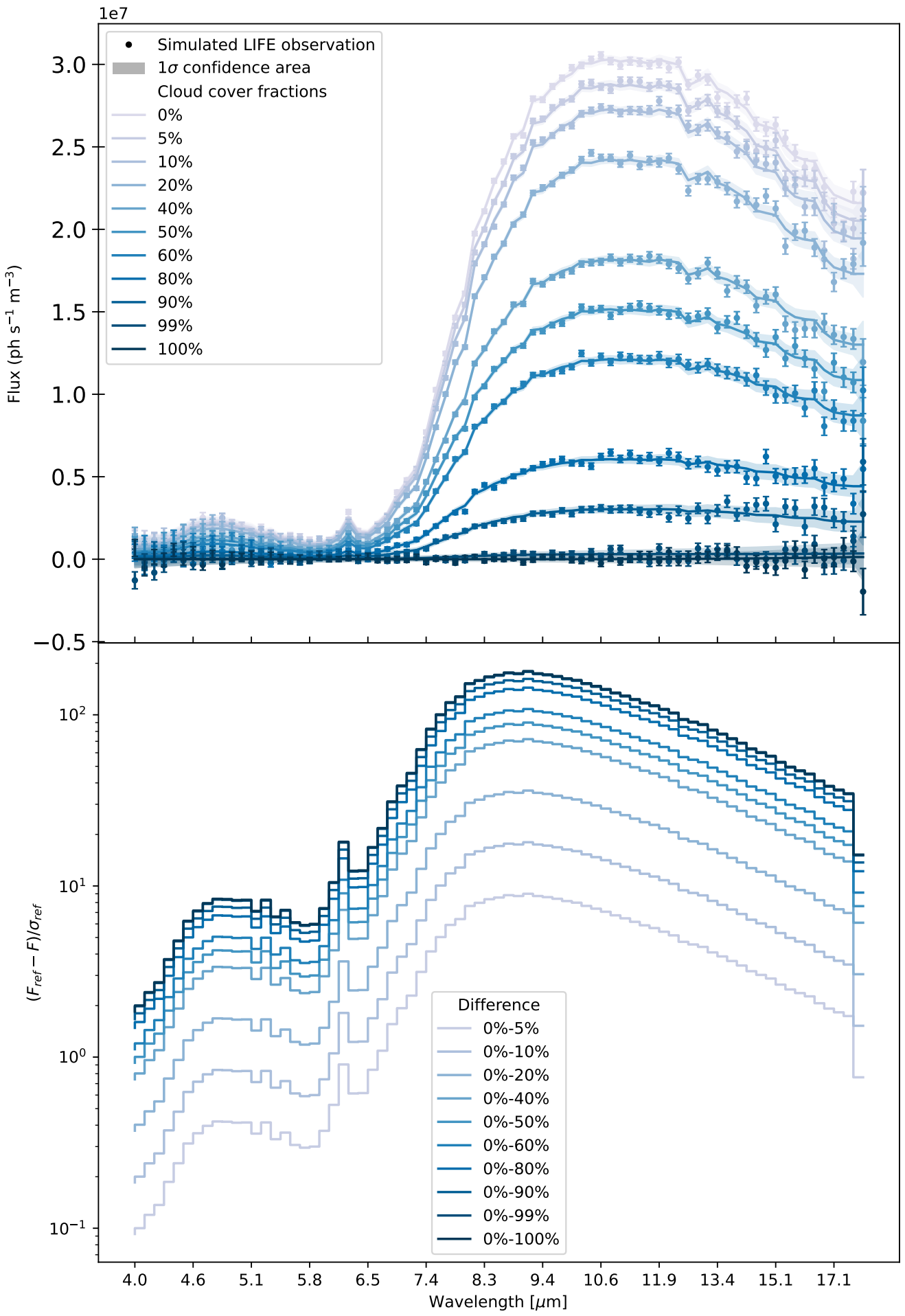

**Figure 11.** The same LIFEsim synthetic observation with an integration time increased to 168h.

LIFE shows a high sensitivity to the cloud cover fraction for wavelengths larger than 7 microns and can meaningfully disentangle scenarios whose cloud fractions differ by 10% given a day of observation, much less given a week. The range from 4 to 7 microns does not allow such a clear distinction, though a case can be made that the possibility remains with one week of observation and even a single day, meaning that the spectral signature of water could be recovered using the two atmospheric windows of water vapor present in this range.

If the atmosphere of Teegarden b is indeed $N_2$-dominated and has a surface pressure lower than 1 bar, the observed flux would yield a good signal to noise ratio and the amplitude of the signal would vary strongly with the amount of $N_2$ present. The resulting spectra might be difficult to interpret, as a higher abundance of $N_2$ may be degenerate with a higher cloud cover fraction. Conversely, if it has at least 1 bar of $N_2$, the degeneracy somewhat disappears because the spectra would have a much weaker dependence on the $N_2$ abundance compared to the cloud cover fraction.

Although there are good reasons to suspect that the Teegarden planets may harbor temperate secondary atmospheres due to their orbits and the low activity of their star, we cannot dismiss the possibility that they have experienced a runaway greenhouse effect in the past. A Venus-like atmosphere would yield distinctive spectra compared to the $N_2$-dominated atmosphere, and LIFE would still retain a similar sensitivity to the cloud cover fraction. Even if water vapor is only a trace gas, the presence of $CO_2$ alone would be visible due to its powerful absorption around 15 $\mu$m. This showcases the versatility of LIFE given its currently conceptualized baseline performance and its wavelength coverage.

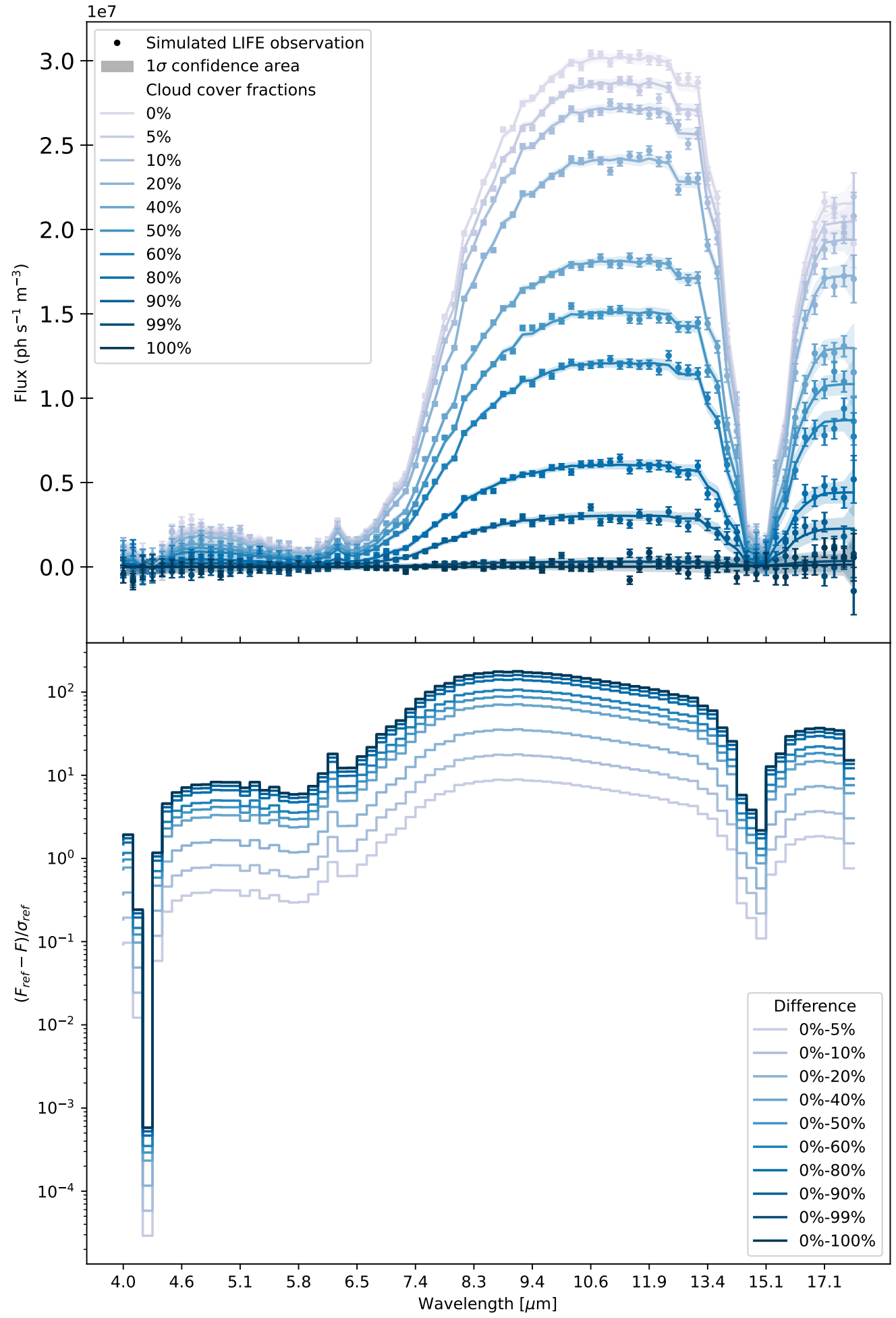

**Figure 12.** The same LIFEsim synthetic observation with an integration time increased to 168h and the addition of 40 Pa of $CO_2$ in the atmosphere.

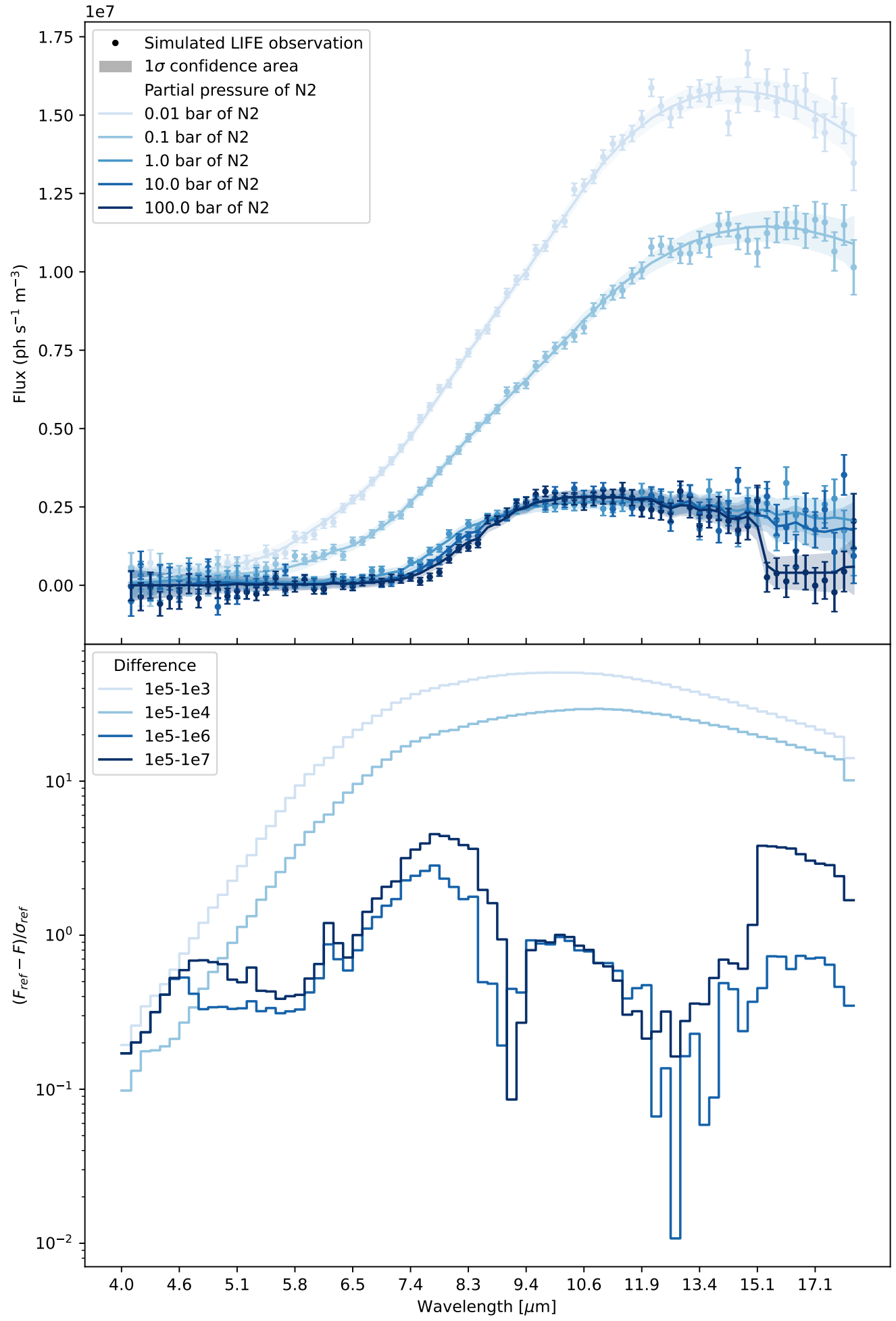

**Figure 13.** LIFEsim synthetic phase curves sweeping over five partial pressures of $N_2$, with a resolution of 50 and an integration time of 24 hours.

## 5. CONCLUSIONS

Teegarden b will likely be one of the first targets of LIFE because of its Earth-like bulk properties, its orbit located in the habitable zone, and the low activity of its star which increases the likelihood that a secondary atmosphere has survived. If there is in fact liquid water and a hydrological cycle in operation, we expect the formation of clouds which will affect observations.

The cloud cover fraction has a significant impact on both the climate and the emission spectra that LIFE would obtain from an observation of Teegarden's Star b. An increased fraction quickly makes the planet reach an equilibrium, and this tendency is linear.

In addition, even a cloud coverage of 90% would yield a detectable signal which is characterizable by LIFE with a high sensitivity, even if the planet has a near-complete water cloud coverage. The detection of an $N_2$-dominated atmosphere could present some degeneracies with cloud cover, but these are reduced for an Earth-like or higher abundance of $N_2$. For a Venus-like scenario, LIFE remains capable of characterizing an opaque $CO_2$-dominated atmosphere and of disentangling cloud cover fractions at least as small as 10%. This is a promising opportunity for future direct imaging observations of temperate planets or Venus analogs that potentially feature clouds, which would be challenging for current instruments that perform transit spectroscopy.

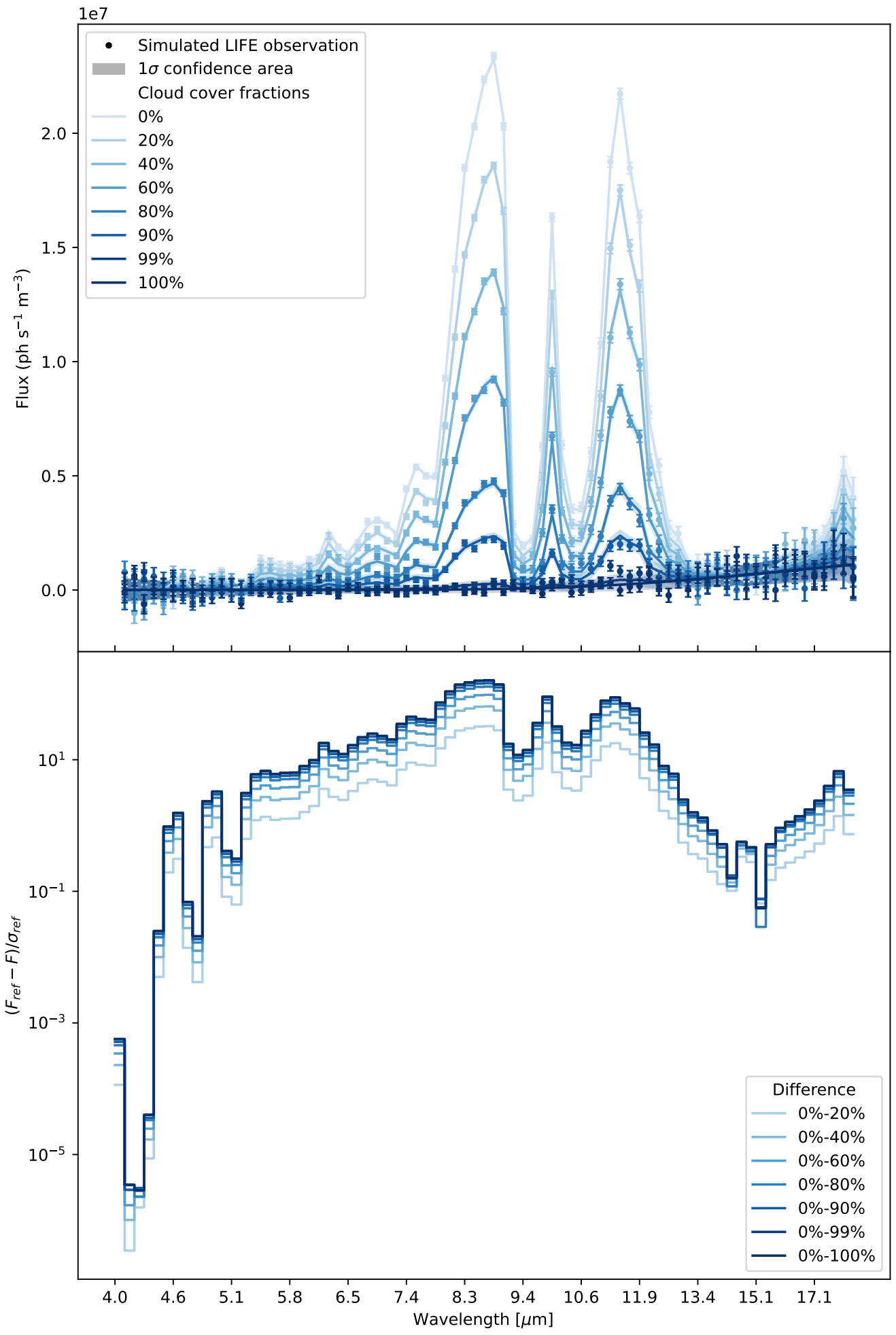

**Figure 14.** Venus-like atmosphere with 3.3 bars of $N_2$ and 89.7 bar of $CO_2$, added to a 7-bar water ocean which is evaporated into the atmosphere. The surface temperature is 740 K. The resolution is 50 and the integration time is 168h.

ACKNOWLEDGMENTS

The authors thank the anonymous referee for helpful comments that improved the manuscript. They also thank Prof. Thorsten Mauritsen and Dr. James Manners for helpful discussions. Some of the computations were enabled by resources provided by the National Academic Infrastructure for Supercomputing in Sweden (NAISS), partially funded by the Swedish Research Council through grant agreement no. 2022-06725. This work was also supported by an interdisciplinary postdoctoral fellowship issued by the Section for Mathematics and Physics at Stockholm University.

*Software:* NUMPY (Harris et al. 2020), SCIPY (Jones et al. 2001–), MATPLOTLIB (Hunter 2007), SEABORN (Waskom et al. 2018), SOCRATES (Edwards & Slingo 1996), LIFESIM (Quanz et al. 2022)

# APPENDIX

Figure 15 shows the spectral sum of the flux contribution function profiles at 250 K and 650 K. It shows both the sum over the full [1, 30000] $cm^{-1}$ spectral range and over the current baseline spectral range of LIFE, [540.54, 2500] $cm^{-1}$.

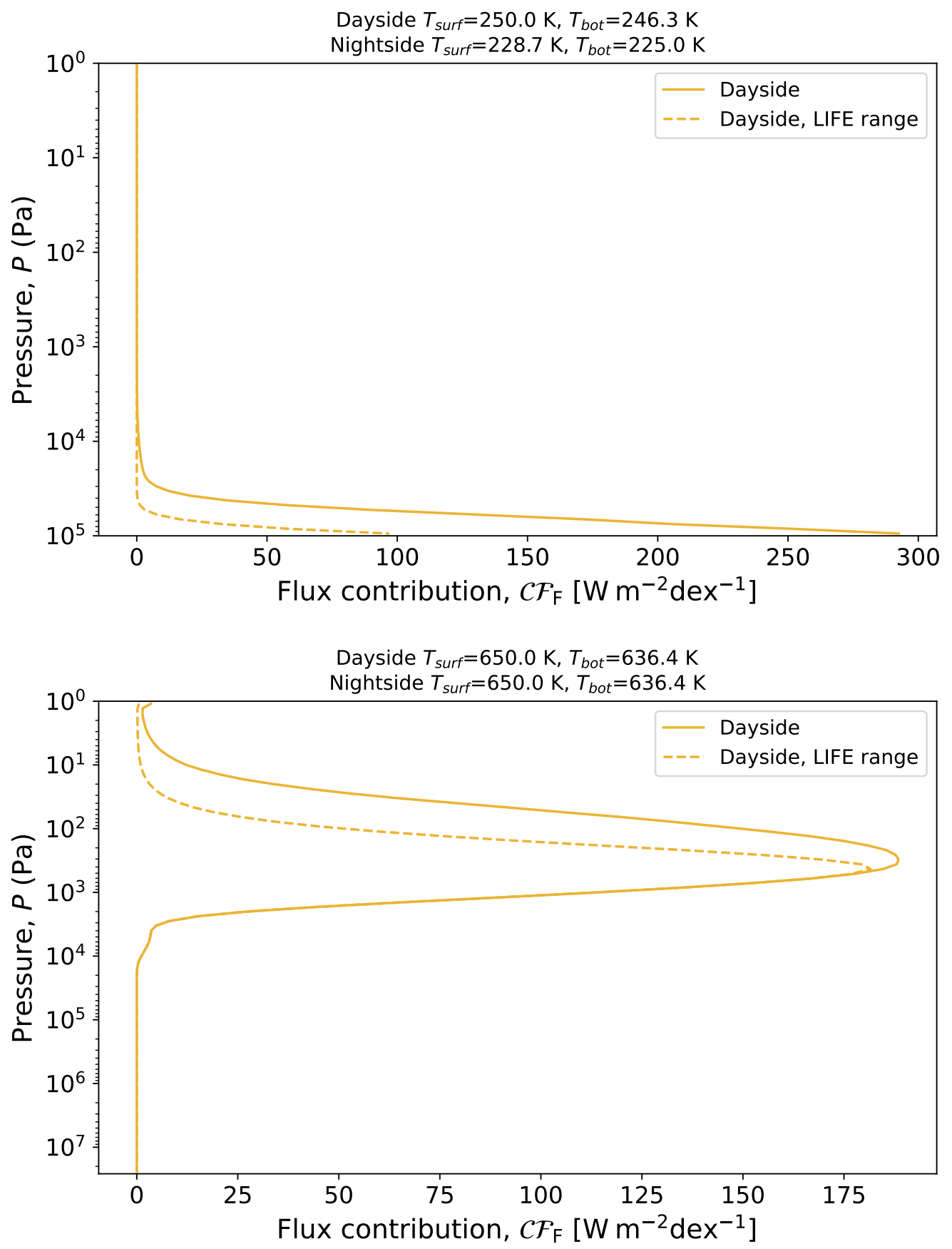

**Figure 15.** Flux contribution function profiles for a surface temperature of **(Top)** 250 K and **(Bottom)** 650 K. Solid lines are the sum over the range [1, 30000] $cm^{-1}$ and the dashes lines are the sum over the spectral range of LIFE, [540.54, 2500] $cm^{-1}$.

Figure 16 shows the vertical variation of the emission spectrum, across the subset of wavenumbers [1, 10000] $cm^{-1}$, assuming a surface temperature of 300 K and a cloud cover fraction of 60%. In this case, the cloud extends from 18 mbar to 82 mbar, so only the red curve representing the emergent flux at the top of atmosphere is affected by absorption from the cloud, as the others are below the cloud base.

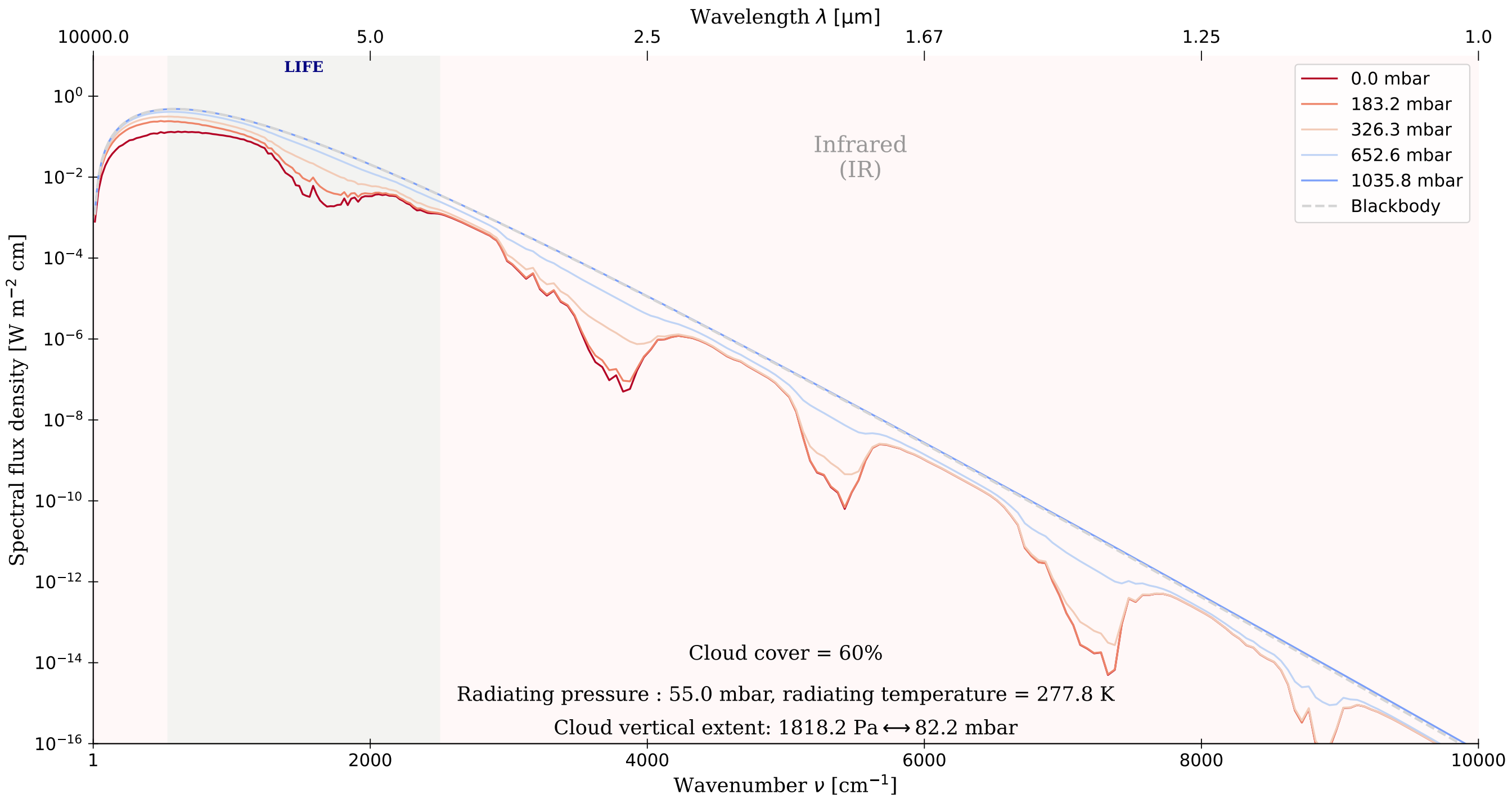

**Figure 16.** Upward radiant fluxes as a function of wavenumber at a surface temperature of 300 K, a cloud cover fraction of 60%, and at four pressure levels spanning the vertical extent of the column. The surface blackbody curve is plotted as reference.